# Automatic Building Code Review: A Case Study

Hanlong Wan[*], Weili Xu, Michael Rosenberg, Jian Zhang, Aysha Siddika[†]

Pacific Northwest National Laboratory, Richland, WA, USA
*Corresponding Author
hanlong.wan@pnnl.gov

This manuscript has been authored by Battelle Memorial Institute under Contract No. DE-AC05-76RL01830 with the US Department of Energy (DOE). The US government retains and the publisher, by accepting the article for publication, acknowledges that the US government retains a nonexclusive, paid-up, irrevocable, worldwide license to publish or reproduce the published form of this manuscript, or allow others to do so, for US government purposes. DOE will provide public access to these results of federally sponsored research in accordance with the DOE Public Access Plan (http://energy.gov/downloads/doe-public-access-plan).

[†] Former intern at Pacific Northwest National Laboratory; currently a student at the University of Wyoming.

# ABSTRACT

Building officials, especially those in resource-constrained or rural jurisdictions, struggle with labor-intensive, error-prone, and costly manual reviews of design documents as projects scale in size and complexity. Widespread adoption of Building Information Modeling (BIM) and Large Language Models (LLMs) has created opportunities for automated code review (ACR) solutions. This study proposes a novel agent-driven framework that integrates BIM-based data extraction with automated verification using both retrieval-augmented generation (RAG) and Model Context Protocol (MCP) agent pipelines. The framework employs LLM-enabled agents to extract geometry, schedules, and system attributes from heterogeneous file types, which are then processed for building code checking via two complementary mechanisms: (i) direct API calls to DOE's COMcheck engine, providing deterministic and audit-ready outputs, and (ii) RAG-based reasoning over rule provisions, allowing flexible interpretation where coverage is incomplete or ambiguous.

The framework was evaluated through case demonstrations, including automated extraction of geometric attributes (e.g., surface area, tilt, and insulation values), parsing of operational schedules, and design validation for lighting allowances under ASHRAE Standard 90.1-2022. Comparative performance tests across multiple large language models showed that Generative Pre-trained Transformer 4 Omni (GPT-4o) achieved the best balance of efficiency and stability, while smaller models exhibited inconsistencies or failures. Results confirm that MCP agent pipelines perform better than RAG reasoning pipelines on rigor and flexibility in workflows. This work advances ACR research by demonstrating a scalable, interoperable, and production-ready approach that bridges BIM with authoritative code review tools.

# NOMENCLATURE

| | |
|---|---|
| ACR | Automated Code Review |
| AEC | Architecture, Engineering, and Construction |
| API | Application Programming Interface |
| ASHRAE | American Society of Heating, Refrigerating and Air-Conditioning Engineers |
| BIM | Building Information Modeling |
| CAD | Computer-Aided Design |
| DOE | U.S. Department of Energy |
| gbXML | Green Building XML |
| GPT | Generative Pre-trained Transformer |
| HVAC | Heating, Ventilation, and Air Conditioning |
| IECC | International Energy Conservation Code |
| IFC | Industry Foundation Classes |
| JSON-RPC | JavaScript Object Notation – Remote Procedure Call |
| LLM | Large Language Model |
| MCP | Model Context Protocol |
| NLP | Natural Language Processing |
| RAG | Retrieval-Augmented Generation |
| ReAct | Reasoning and Actions |

# 1     INTRODUCTION

In order for American businesses, building owners, and tenants in commercial and multifamily buildings to realize the cost savings from building codes such as ANSI/ASHRAE/IES Standard 90.1 – Energy Standard for Buildings Except Low-Rise Residential Buildings (ASHRAE 90.1) and the International Energy Conservation Code (IECC), it is important to verify the installation of the building envelope, service water heating, lighting, Heating, Ventilation, and Air-Conditioning (HVAC) systems, as well as their interactions. Building officials and code reviewers struggle with labor-intensive, error-prone, and costly manual reviews of design documents, especially as Building Information Models (BIM) grow in size and complexity. To address these challenges, Automated Code Review (ACR) has emerged as a critical enabler for streamlining design verification, assisting automatic building code review, and accelerating approval processes.

## 1.1    Key Trends in Building Plan Verification

With the adoption of digital design tools such as BIM, building design verification workflow is ripe for digitization. Manual checking, traditionally prone to errors and inefficiencies, is now increasingly replaced by automated systems that simplify workflows and accelerate the review process. For example, Xue et al. highlighted the value of semi-automated methods by converting code tables into structured rules, particularly addressing overlooked regulatory requirements stored in spreadsheets [1]. Similarly, Wu & Zhang validated inferable concepts such as fire safety egresses using rule-based logic, achieving major reductions in the time required for code checks [2].

Computational techniques underpinning ACR continue to evolve, with approaches ranging from deterministic rule-based hardcoding to ontology-driven reasoning and, more recently, natural language processing (NLP) and large language models (LLMs).

### 1.1.1    Rule-based hardcoding

Rule-based hardcoding, which encodes regulatory provisions directly into scripts or decision rules, has been a foundational attribute of the early ACR process due to its deterministic nature and accuracy. For instance, Nawari utilized fuzzy logic alongside hardcoded frameworks optimized for the Industry Foundation Classes (IFC) schema, enabling qualitative reasoning and handling ambiguous rules [3]. Zheng et al. presented a hard-coded, LiDAR (Light Detection and Ranging)-driven methodology using segmentation and filtering techniques in MATLAB to automate millimeter-level geometric quality assessment of railings, cutting inspection time and labor while ensuring safety checks [4]. Reinhardt & Mathews similarly employed hardcoded scripting tools for design verification execution in BIM, combining deterministic algorithms with visual programming [5].

### 1.1.2    Ontology-based reasoning

Ontology-based reasoning leverages structured knowledge representations to capture building concepts, relationships, and constraints, enabling automated design checking by aligning BIM data with regulatory requirements. Such systems are more scalable and flexible than rigid hardcoding approaches. Jiang et al. presented a multi-ontology model reconciling IFC-compatible BIM metadata with regulatory clauses [6], while Ma et al. demonstrated how SPARQL (SPARQL Protocol and RDF Query Language) and SWRL (Semantic Web Rule Language) inference frameworks improve consistency across projects [7]. Ontology-based systems also increasingly integrate knowledge graphs, as shown by Zhu et al., to dynamically validate certain conditions such as fire exits and egress paths [8].

### 1.1.3 Natural Language Processing and Large Language Model

NLP, a subfield of artificial intelligence, enables machines to analyze, interpret, and generate human language [9]. Recent advances in LLMs, such as GPT-3 and GPT-4, have significantly expanded the capabilities of ACR systems by leveraging transformer architectures trained on massive corpora to interpret and translate regulatory language into actionable logic [10,11]. For instance, Madireddy et al. employed GPT-4, Claude, Gemini, and Llama to generate Python scripts for Revit-based checks, iteratively refining prompts and achieving measurable improvements in success rates [12]. Similarly, Chen et al. combined Bidirectional Encoder Representations from Transformers, GPT-4, and Convolutional Neural Network for Text models to enhance rule classification and extract patterns from complex regulatory documents [13].

### 1.1.4 Challenges

Despite technological progress, current ACR systems still face challenges with incomplete or inconsistent BIM metadata, fragmented adoption of different code editions across jurisdictions with local amendments, and the lack of context for interpreting qualitative or performance-based rules. Peng & Liu noted limitations in IFC data that restrict semantic expansion for automated rules [14], while Bloch & Sacks highlighted the challenge of missing or implicit metadata in BIM models [15]. Handling complex regulatory provisions also remains a bottleneck. For instance, Doukari et al. emphasized that encoding performance-based rules often requires manual intervention [16], and Chen & Jiang pointed out the difficulty of applying NLP methods to semantically vague fire safety regulations [17]. Moreover, systemic resistance—such as limited BIM infrastructure among small firms and skepticism from Building Officials—further hinders widespread adoption [18].

These challenges highlight the need for next-generation ACR systems that can operate across fragmented data sources, adapt to evolving requirements, and seamlessly interface with existing industry tools. ACR solutions could move beyond static rule encodings to embrace flexible, intelligent frameworks capable of handling incomplete BIM data, heterogeneous document formats, and performance-based design file provisions.

## 1.2 COMcheck

COMcheck [19] is an automatic document-review software tool developed by the U.S. Department of Energy to assist architects, engineers, and building professionals in verifying that commercial and high-rise residential projects meet certain standards, such as the IECC and ASHRAE Standard 90.1, as well as various state-specific codes. Originally delivered as a desktop application, COMcheck has since evolved into a web-based system, COMcheck-Web, which eliminates the need for local installation and provides enhanced usability, performance, and support for newer code versions (e.g., IECC 2024, ASHRAE 90.1-2022). More recently, the COMcheck Application Programming Interface (API) was introduced. Built on modern serverless cloud infrastructure, the API enables dynamic scalability and seamless integration across verification workflows. It also makes it possible to embed verification functionality directly into web applications, facilitating data entry, automated reporting, and incorporation into broader building-design workflows.

Although other check tools exist, such as California Building Energy Code Compliance – Commercial [20], this paper focuses on COMcheck because its recently released API uniquely supports the type of LLM-based agent call explored in this work. COMcheck therefore serves as a representative example of how certain software can evolve to support more automated, API-driven checking approaches.

## 1.3 AI Agents and the Model Context Protocol

LLMs are powerful systems trained on vast text corpora, capable of few-shot learning (making accurate predictions after seeing only a few examples) and complex instruction following, especially when fine-tuned with methods such as reinforcement learning from human feedback. Building on LLM capabilities, AI agents are autonomous systems that decompose high-level tasks into sequences of reasoning and tool interactions. A prominent agentic framework is ReAct (Reason + Act) [21], which interleaves natural language reasoning with executable actions—enabling LLMs to iteratively plan, act on external APIs or environments, observe results, and update their internal reasoning. This synergy between "thought" and "action" has been empirically shown to improve interpretability, reduce hallucination, and enhance performance on tasks like question answering and interactive decision-making.

The Model Context Protocol (MCP) [22] provides an open, standardized mechanism built on JSON RPC 2.0 for seamless interoperability between LLM based agents and external tools or data sources. MCP abstracts integration complexities by allowing agents (as clients) to discover and invoke functionality from MCP servers (for example, databases, APIs, file systems) in a secure, modular fashion, addressing the "N×M" integration challenge with a unified interface akin to a universal connector for AI tooling. Since its release, MCP has gained traction, with notable adoption by OpenAI, Google DeepMind, and enterprise agents, laying the groundwork for scalable, interoperable, tool aware LLM systems.

This paper advances the field by proposing a novel integration of BIM with AI-driven agents for compliance automation. Specifically, the framework leverages existing BIM parsing tools such as lxml [23] to automatically extract structured building data from gbXML [24] format file, which is then processed through agent-based orchestration. The agents are capable of invoking external compliance tools such as the COMcheck API to perform regulation-specific validation without manual intervention. Unlike retrieval-based approaches, which may introduce ambiguity or errors in interpreting code clauses, this agent-driven pipeline ensures accuracy, scalability, and labor savings by directly interfacing with authoritative compliance engines. The contributions of this work lie in demonstrating (1) a tightly integrated workflow linking BIM data to automated compliance checks, (2) the application of multi-agent AI systems to coordinate BIM parsing, rule interpretation, and tool invocation, and (3) an evaluation of efficiency and accuracy improvements compared to traditional manual or RAG-based methods. Together, these innovations highlight a path toward practical, production-ready ACR that bridge BIM environments and verification platforms.

# 2 METHODOLOGIES

## 2.1 Workflow

The proposed methodology for automated compliance checking follows a modular workflow consisting of two primary stages: data extraction and compliance checking, as shown in Figure 1. At the outset, the system ingests architectural design plans, typically available in BIM, Computer-Aided Design (CAD), and Portable Document Format (PDF) formats. A data extraction pipeline processes these heterogeneous sources to derive essential building information, such as room dimensions, wall properties, surface tilt, occupancy schedules, lighting schedules, and HVAC system types. These extracted data are consolidated into a unified representation describing building types, geometry, operating schedules, and mechanical system characteristics.

In the subsequent stage, a compliance checking module evaluates the extracted information against relevant building documentations. This is achieved through two complementary mechanisms: (i) automated queries to existing compliance check tools (e.g., COMcheck API) and (ii) LLM-based reasoning with embedded document provisions. The workflow is designed to support interoperability, minimize manual data entry, and enable scalable evaluation across diverse design iterations.

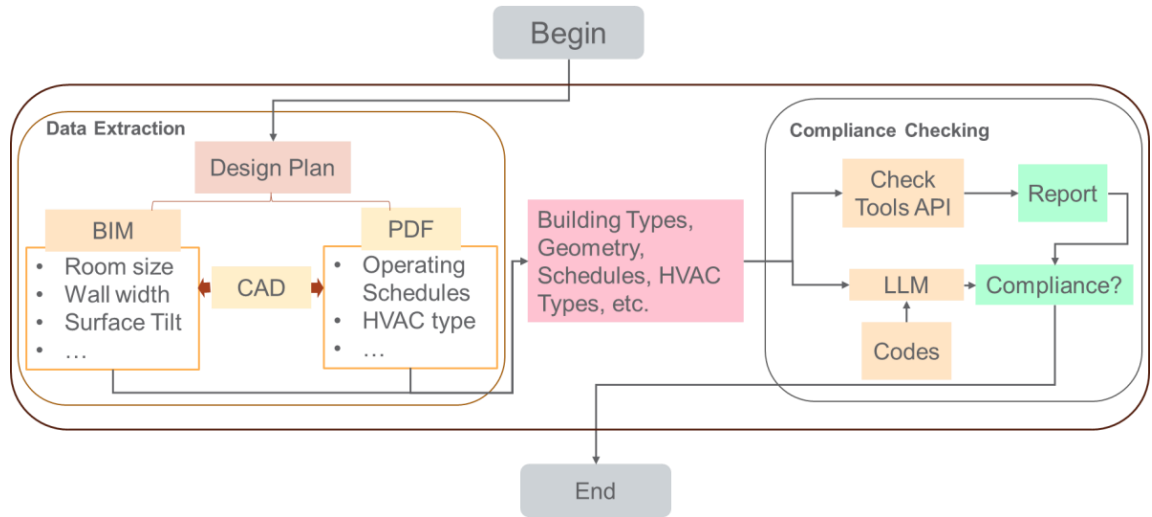

**Figure 1 Workflow of the ACR from BIM and Architecture Design File**

## 2.2 Data extraction

The data extraction stage is responsible for transforming unstructured or semi-structured design documents into machine readable attributes required for compliance evaluation. BIM files provide granular geometric and spatial details, including room sizes, surface orientations, and wall configurations. These geometric data are supplemented with operational and system-level information such as occupancy schedules and HVAC system types, commonly available in PDF based design specifications. When CAD files are present, they act as an intermediary, ensuring alignment between BIM geometry and PDF based metadata.

By integrating these heterogeneous sources, the methodology generates a consistent dataset that captures both physical and operational attributes of the building. This integrated dataset forms the basis for reliable compliance evaluation and ensures that both geometric fidelity and operational context are preserved.

## 2.3 Compliance checking

Once the necessary building attributes are compiled, the methodology advances to the compliance checking stage. Two complementary approaches are adopted. First, structured data are formatted into standardized inputs for the COMcheck API, a compliance tool developed by the U.S. Department of Energy (DOE) to verify adherence to ASHRAE Standard 90.1 and the IECC. The API returns machine-readable reports indicating pass/fail outcomes and highlighting specific deficiencies in envelope, lighting, or HVAC systems.

Second, a retrieval-augmented generation (RAG) pipeline integrates LLM-based reasoning with codified text. The LLM retrieves relevant provisions from ASHRAE or IECC standard files and interprets building data in this context to determine compliance status. This dual approach allows for both formal verification via established compliance tools and flexible reasoning for cases where automated tools may not provide sufficient coverage.

Together, these mechanisms deliver a comprehensive compliance evaluation, producing structured reports that can guide design revisions and support iterative decision-making.

# 3 RESULTS

## 3.1 Data Extraction

### 3.1.1 BIM File Extraction

The first demonstration of the proposed methodology focused on geometry data extraction from BIM files in gbXML format. Figure 2 illustrates the interactive pipeline in which user queries are translated into geometry attributes through an LLM agent. When a query such as "What is the size of the surface?" is posed, the agent interprets the request, generates the appropriate tool call (e.g., get_surface_area), and retrieves results directly from the gbXML data.

For example, querying the ceiling surface ceiling_unit1_Reversed yielded a calculated area of 110.41 m², while complementary tool calls extracted additional parameters such as surface tilt and thermal resistance (R-value). The agentic workflow ensures that each query is automatically mapped to structured tool calls, thereby eliminating manual inspection of gbXML schemas and accelerating the retrieval of building attributes. Results are returned in a human-readable format through the LLM interface, while also being retained in structured JSON for downstream compliance checking.

This demonstration confirms that the framework can accurately extract critical geometric properties—surface area, tilt, and insulation levels—from BIM files. The ability to query geometry at this granularity enables precise alignment of design data with code compliance requirements.

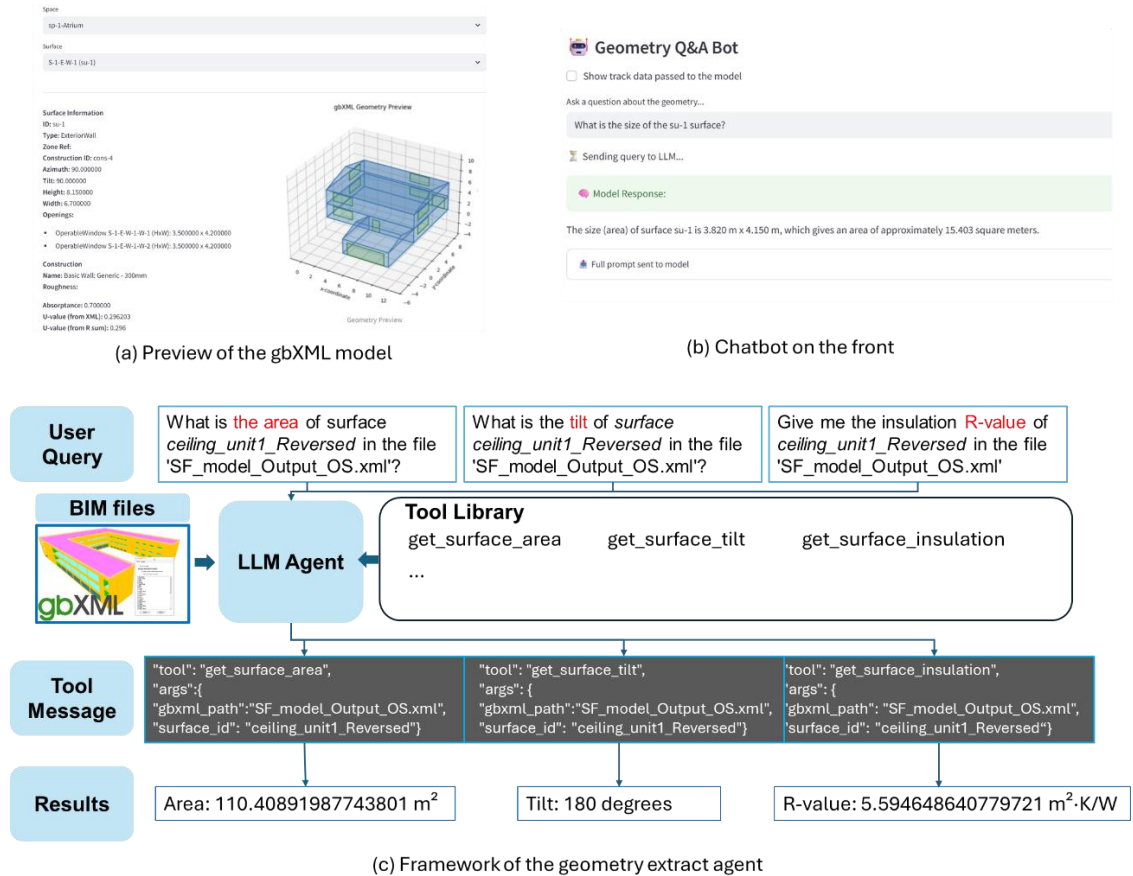

**Figure 2 Demonstration of BIM geometry data extraction**

### 3.1.2 Design File Extraction

Beyond geometric properties, many essential parameters for compliance checking reside in architectural design documents not directly available in BIM. The methodology was further applied to extract operational and system-level data including lighting power densities, occupancy and operating schedules, and HVAC system types.

The extraction pipeline integrates document parsing with LLM-based reasoning to identify relevant attributes within semi-structured or unstructured text. For instance, lighting schedules embedded in design specifications were successfully parsed into structured tabular formats representing weekday and weekend operating hours. Similarly, HVAC system descriptions were mapped to categorical variables (e.g., cooling type, heating fuel, ventilation strategy) consistent with compliance tool requirements.

A case study was performed to demonstrate robustness against formatting issues as shown in Figure 3. Here, design files were converted into figures, and optical character recognition (OCR) was applied to extract text reliably, overcoming font inconsistencies and layout problems that would otherwise block correct information retrieval. The results indicate that this approach can transform diverse architectural files into standardized, machine-readable inputs. When combined with the geometry extraction, these outputs provide a complete representation of building attributes necessary for automated compliance evaluation.

(a) Chatbot on the front

(b) Output prompts

**Figure 3 Example of lighting schedule extraction from design files**

## 3.2 Compliance Check

The final stage of the methodology demonstrates how extracted building attributes can be integrated with compliance engines for automated verification. Two complementary approaches are considered: 1) RAG-based QA, where knowledge from a specific code file is retrieved to answer user queries directly through an LLM and 2) COMcheck-integrated agents, which invoke existing hard rule–based tools through API calls, while leveraging LLM reasoning for query interpretation.

### 3.2.1 RAG-based QA

In this approach, code documents are indexed using RAG. When a user submits a compliance-related query, the system retrieves the most relevant passages from the code and prompts the LLM to generate an answer grounded in the retrieved text. The process is very similar to what has been reported in section 3.1.2 to extract information from a design file. This work was reported in [25].

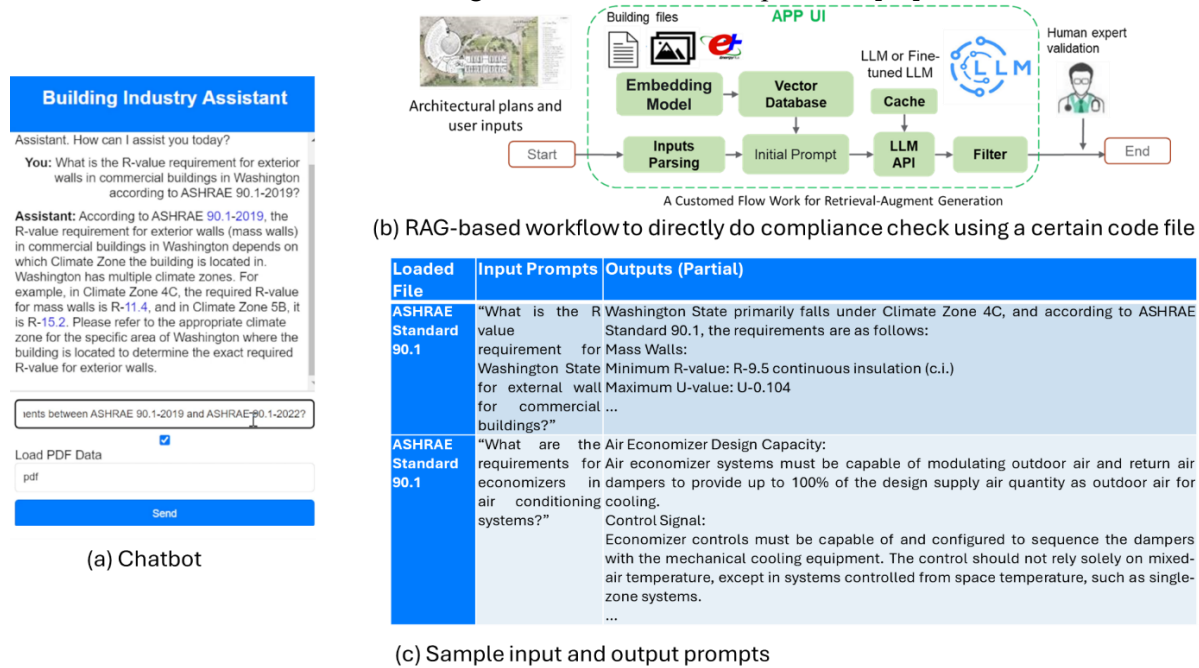

**Figure 4 An example of RAG-based compliance checking [25]**

### 3.2.2 COMcheck-Integrated Agent

Figure 5 illustrates an end-to-end example where a user query, "What is the lighting power allowance for a 500-square-meter bank according to ASHRAE Standard 90.1-2022?", is processed through the compliance agent.

The workflow begins with the agent parsing the query into standardized inputs—identifying the floor area, building type (bank), and applicable code version (ASHRAE 90.1-2022). These inputs are translated into API calls referencing the COMcheck tool library. During execution, the agent employs chain-of-thought reasoning to resolve ambiguities, such as unit conversions[‡] (e.g., square meters to square feet) and validation of code versions. This iterative reasoning ensures compatibility between user inputs and COMcheck API specifications.

The API returns a structured result, which the agent interprets and reformats into a final output. In this example, the lighting power allowance was successfully calculated as 3,019 W, consistent with the requirements of ASHRAE 90.1-2022 for the specified building type and size. The result was simultaneously presented in a user-friendly natural language format and stored as machine-readable JSON, enabling both interpretability and downstream integration.

---

[‡] The requirements in the IP and SI versions of Standard 90.1 are not always exact translations. For instance, 25,000 ft² in the IP edition corresponds to 2,300 m² in the SI edition. The AI currently performs only direct unit conversions, which can introduce slight discrepancies. Addressing this nuance is a potential area for future study.

This demonstration confirms that the combined use of the COMcheck API and LLM-based reasoning can automate the traditionally manual process of compliance verification. By bridging between unstructured user queries and standardized compliance engines, the system supports accurate, transparent, and repeatable assessments of building code adherence.

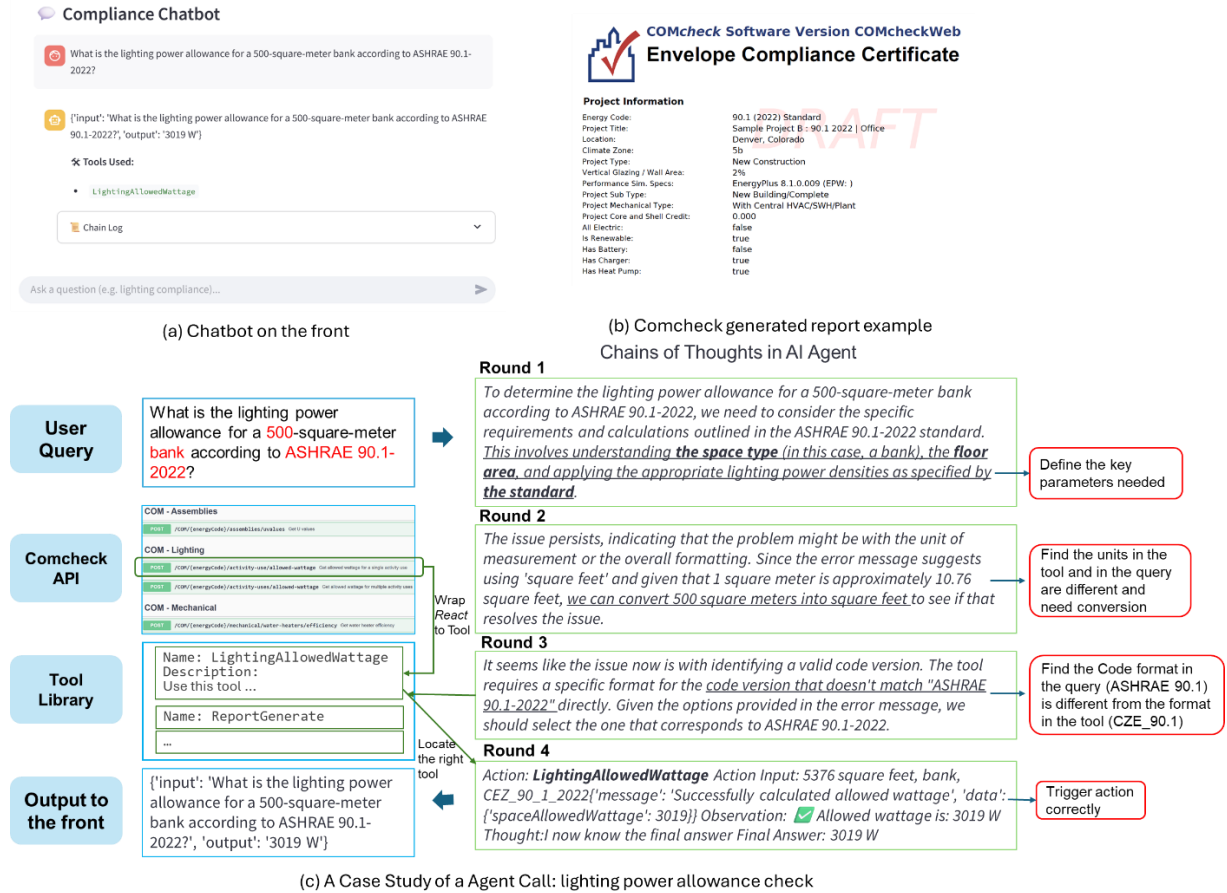

Figure 5 Demonstration of compliance check based on COMcheck

## 4   DISCUSSION

### 4.1   Model performance

Figure 6 presents the response time and token consumption of different large language models across two representative prompts as shown in Appendix 1. Clear performance differences emerge among the models. GPT-4o exhibited the fastest average response time and the lowest variance, reflecting both efficiency and stability. In contrast, o4-mini showed the slowest average response and the widest variability, indicating inconsistency under comparable workloads. The Claude models achieved intermediate response times with relatively narrow spreads, demonstrating stable behavior across runs. The o3-mini model failed to generate valid outputs for the tested prompts and was therefore excluded from the comparison.

Token usage followed a similar pattern. GPT-4o consumed fewer tokens on average compared to o4-mini, which displayed wide variability with some runs exceeding 1,000 tokens. The Claude models again showed moderate and stable token usage across both prompts, aligning with their consistent response times.

Overall, GPT-4o achieved the best balance of speed and token efficiency, delivering valid responses quickly while consuming fewer tokens than most other models. The Claude models provided reliable and consistent performance, while o4-mini exhibited unstable behavior.

Temperature settings further influenced GPT-4o's token usage, as shown in Figure 6 (b). At lower temperatures (≤0.7), token counts for both prompts remained compact and stable, indicating controlled and efficient generation. At higher temperatures (≥0.9), token usage increased substantially, with maximum values spiking above 1,000 tokens. This reflects a trade-off: higher temperatures encouraged more verbose and variable outputs, while lower temperatures yielded more concise and predictable responses.

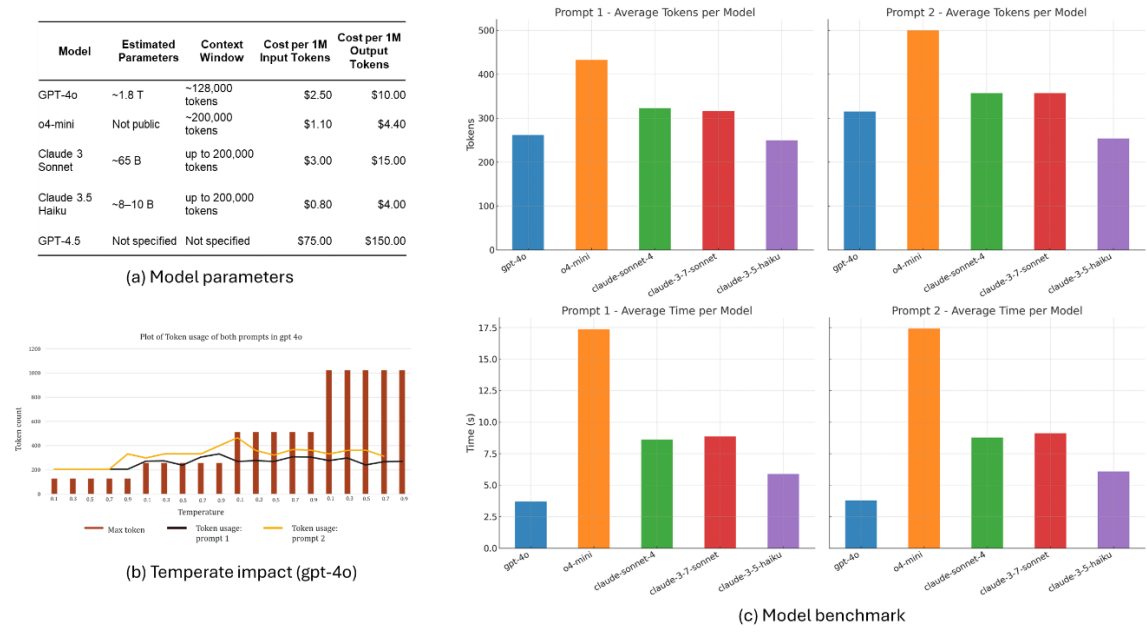

**Figure 6** Different Model Performance on RAG based Check (a) model parameters (b) temperature setting impact on model performance (c) model benchmark

## 4.2 Pipeline comparisons

The compliance checking methodology explored in this study was implemented through two complementary pipelines: a Retrieval-Augmented Generation (RAG) approach and an MCP agentic pipeline interfacing with the COMcheck API (**Error! Reference source not found.**). Both aim to automate the traditionally manual process of code verification, yet they differ in reliability, flexibility, and determinism.

The RAG pipeline draws on a database of code texts such as ASHRAE Standard 90.1 and IECC provisions. When queried, the system retrieves potentially relevant clauses and synthesizes a response via a large language model. This setup offers interpretability by providing both natural language explanations and code excerpts, and adaptability since new documents can be ingested without structural changes to the workflow. However, the results shown in Figure 7 highlight a key limitation: variability. For the same prompt about lighting power allowance, different LLMs returned inconsistent answers (for example, 5500 W, 7585 W, 5400 W, 3019 W), depending on which table (Table 9.5.1 versus Appendix G3.8) the model associated with the query. Interestingly, larger models such as GPT 4 and GPT 5 often defaulted to Appendix G values rather than design case values, likely because more open-source data and online examples emphasize performance rating cases rather than prescriptive design allowances. Moreover, when web search was enabled, GPT 4 and GPT 5 occasionally produced the correct zero-shot answers, perhaps due to "knowledge leakage" where portions of the standard became exposed in public datasets during training, but this success was inconsistent. These findings underscore that while RAG and zero-shot pipelines can sometimes provide accurate results, they are not yet dependable for regulatory implementation.

By contrast, the MCP agent pipeline connects directly with the DOE COMcheck API. Here, building attributes are structured into standardized inputs, and the API evaluates them deterministically against codified rules. This method yields machine-readable, audit-ready outputs that align with review procedures. In the example task, the agent consistently produced the correct 3019 W allowance, avoiding the variability seen with RAG or LLM-only approaches. However, this determinism comes with a narrower scope: the pipeline can only check what COMcheck has encoded, limiting flexibility for unmodeled or qualitative rules.

Taken together, these results illustrate a fundamental trade-off. The RAG pipeline offers breadth, interpretability, and adaptability, useful for early-stage "what-if" analysis or when human designers need contextual reasoning. The agentic pipeline, by contrast, guarantees rigor and reproducibility, making it suitable for certification, documentation, and enforcement. A hybrid workflow, in which RAG provides interpretive reasoning and agentic pipelines deliver formal validation, may therefore be the most effective strategy for scalable, transparent, and trustworthy implementation automation.

Figure 7 Pipeline Comparison for Query of "What is the lighting power allowance for a 500-square-meter bank according to ASHRAE 90.1-2022": (a) Direct LLM Query, (b) RAG with ASHRAE 90.1-2022, and (c) Agentic Call via COMcheck API

### 4.3 Possible future work

The results of this study demonstrate the feasibility of linking BIM data extraction with automated sample building code review and verification through agent-based orchestration and existing check tools. Nonetheless, several research directions remain open for advancing the reliability, scalability, and adoption of ACR.

First, future work can expand beyond building energy topics to include the whole construction lifecycle, even for projects without comprehensive BIM data such as small-scale residential buildings or renovations. In such cases without BIM, data may come from manual inputs, sensor logs, or minimal digital records. To support these contexts, ACR must accommodate both structured BIM outputs and user-entered parameters or simplified metadata, enabling coverage across the spectrum of project types, from full-scale design models to hand-drawn or legacy documentation.

Second, future research could focus on developing review tools that can track branching updates across multiple sources, including evolving BIM exchange formats such as IFC and gbXML, jurisdiction- or state-specific amendments, and new editions of regulatory codes. Ontology-driven methods combined with AI techniques can be used to construct knowledge graphs that encode both prescriptive and performance-based provisions. With version-tracking capabilities similar to those used in software repositories, such tools would make it easier to manage customized updates, review projects under the correct regulatory context, and maintain interoperability as codes continue to evolve.

Third, accuracy and usability in data extraction require further development. Real-world projects often rely on incomplete, inconsistent, or even scanned documentation. Extending the pipeline with computer vision, OCR, and multimodal LLMs could enable resilient extraction from varied file types, including raster drawings, PDF tables, and handwritten notes. A practical interim approach is to adopt a human-in-the-loop semi-automation process, where the system generates a list of missing or uncertain information that requires manual input. As AI and computer vision techniques continue to advance, such interventions are expected to decrease over time, making the extraction process increasingly seamless.

Fourth, real-time and user-centric deployment represents a promising trajectory. Embedding ACR agents into interactive design environments (like Revit or SketchUp) would provide immediate feedback, reducing late-stage revisions. Coupling ACR with IoT-enabled monitoring also offers opportunities to extend verification into construction and operation phases. However, leveraging such real-time monitoring requires addressing practical constraints: while technologies like Control Strainer enable continuous implementation verification using Building Automation System data streams [26], widespread adoption remains limited by security, interoperability, and device coverage challenges.

Fifth, seamless data transfer between different agents or pipeline components remains a critical limitation. Current systems often operate in silos, such as BIM extraction agents, reasoning modules, and rule-checking engines, without robust protocols for interoperability, version control, or state synchronization. Addressing this will require standardized interfaces, shared schemas, and agent coordination frameworks to enable modular, scalable ACR ecosystems.

Finally, benchmarking and standardization remain essential. ACR evaluations often lack consistent metrics, which limits comparability across systems. Establishing shared benchmarks that address accuracy, latency, token efficiency, cost, and performance across different modalities would enable more systematic comparison of LLM-based and traditional rule-based methods. Simultaneously, collaboration with code development bodies, software vendors, and industry stakeholders will be essential to build consensus on standardized data schemas, APIs, and certification protocols.

## 5 CONCLUSION

This study presented a novel framework for ACR that integrates BIM data extraction with agent-orchestrated workflows and existing check tool engines. By combining geometry and system attribute extraction from BIM and architectural design files with both RAG reasoning and MCP-based agent calls to the COMcheck API, the approach demonstrates how LLM-enabled agents can bridge unstructured design information and authoritative building code verification.

The results highlight several contributions. First, the framework successfully automated geometry and schedule extraction from heterogeneous file types, producing structured inputs consistent with check tool requirements. Second, the MCP agent pipeline delivered deterministic and audit-ready outcomes, while the RAG pipeline provided flexible interpretability for ambiguous cases. Third, comparative performance tests across different large language models underscored trade-offs between latency, token usage, and reliability, with GPT-4o achieving the best balance of efficiency and stability.

Collectively, these findings confirm that ACR can be extended beyond static rule-based scripts to more adaptive, agent-driven architectures capable of supporting diverse implementation scenarios. The proposed workflow reduces manual data entry, increases transparency, and accelerates the verification process, which can lower design and review costs for businesses, building owners, and tenants, while also easing the workload of under-resourced building departments, particularly those in rural areas. This positions it as a scalable solution for streamlining real-world building design and code enforcement.

Future work should focus on extending the framework to additional code domains such as fire safety, accessibility, and zoning; enhancing multimodal data extraction from incomplete or scanned documents; and developing standardized benchmarks for performance evaluation. By continuing to align technical innovations with both industry cost savings and public-sector capacity constraints, future research can maximize the impact of AI-enabled review systems across the building ecosystem. With continued research and collaboration among researchers, software vendors, and code development bodies, the integration of AI agents with BIM has the potential to deliver reliable, real-time, and industry-wide checking and review automation.

## AUTHORSHIP CONTRIBUTION STATEMENT

**Hanlong Wan**: Methodology, Validation, Formal analysis, Software, Writing – Original Draft, Visualization, Investigation. **Weili Xu**: Methodology, Writing – Original Draft – COMcheck & MCP. Project administration. **Michael Rosenberg**: Supervision, Funding acquisition, Writing – Review & Editing. Project administration. **Jian Zhang**: Conceptualization, Methodology, Writing – Review & Editing. **Aysha Siddika**: Writing – Original Draft – Literature Review.

## DECLARATION OF COMPETING INTEREST

The authors declare that they have no known competing financial interests or personal relationships that could have appeared to influence the work reported in this paper.

## DECLARATION OF GENERATIVE AI AND AI-ASSISTED TECHNOLOGIES IN THE WRITING PROCESS

During the preparation of this work, the authors used GPT4o in order to improve language. After using this tool/service, the authors reviewed and edited the content as needed and take full responsibility for the content of the publication.

## ACKNOWLEDGMENTS

The authors would like to express their gratitude for the support of the U.S. Department of Energy's Office of Energy Efficiency and Renewable Energy (EERE) through Battelle Memorial Institute under Contract No. DE-AC05-76RL01830. The authors would also like to acknowledge Jeremy Williams and Perry Christopher – Building Energy Code Program, DOE Building Technologies Office.

# APPENDIX I PROMPTS

### a. Model performance

**Prompt 1**
What is the minimum U-factor required for a doorway in Climate Zone 5 according to ASHRAE 90.1-2022?

**Prompt 2**
Which IFC entity type is used to represent a building envelope wall for energy code compliance checks?

### b. Data extraction

**System prompt**
You are an expert assistant helping users understand building data.
Use the provided information to answer questions.

### c. Agent

**System prompt**
"role": "system",
"content": (
    "You are an agent tool caller. Follow this rule strictly:\n"
    "1. If a tool is needed, only respond with:\n"
    "   Action: <tool name>\n"
    "   Action Input: <input text>\n"
    "2. Do NOT include Final Answer until after tool output.\n"
    "3. NEVER return both Action and Final Answer in the same response.\n"
    "4. NEVER invent tool results — wait for actual tool output."